\documentclass[aps,prb,twocolumn,groupedaddress]{revtex4}
\usepackage{graphicx}

\begin{document}


\title{Relativistic electron energy loss and induced radiation emission 
in two-dimensional metallic photonic crystals I: 
formalism and surface plasmon polariton}

\author{Tetsuyuki Ochiai and Kazuo Ohtaka}
\affiliation{Center for Frontier Science, Chiba University, 
Chiba 263-8522, Japan}


\date{\today}

\begin{abstract}
A fully relativistic description of the electron energy loss and the induced 
radiation emission in arbitrary arrays 
of non-overlapping metallic cylinders is presented in terms of the multiple 
scattering method on the basis of  vector cylindrical waves.  
Numerical analysis is  given for dilute and dense arrays of 
Aluminum cylinders with a nanoscale diameter. 
The results of the electron energy loss spectrum are 
 well correlated with the dispersion relation of coupled surface 
plasmon polaritons, and can be 
interpreted with an effective medium approximation when 
the electron runs inside the arrays. 
In addition, the cavity modes localized in the grooves between the cylinders 
can affect strongly the electron energy loss spectrum.    
\end{abstract}


\maketitle


\section{Introduction}

Recently, much interest has been attracted in optical properties of 
composite materials such as cluster of metallic nano-particles
\cite{nano}, 
photonic crystal\cite{Joan,Sakoda}, and left-handed material
\cite{Veselago,Pendry1,Shelby}. 
 They have the potential ability to enhance various optical processes 
with their  rich spectrum, and will be a key component of 
future optoelectronics devices.  
So far, their properties have been investigated mainly with far-field 
optical equipment. However, near-field responses of them are of great 
importance, because much information can be attained from the evanescent 
light involved in composite materials. 

The Electron energy loss (EEL) spectroscopy in a scanning transmission 
electron microscope is a useful tool to 
investigate both surface and bulk excitations of samples.
In a typical EEL experiment the kinetic energy of electron 
is on the order of 100 [keV], and a low-energy part of the EEL spectrum 
is related to a collective excitation such as surface plasmon. 
Thus, the spectrum is well interpreted 
in terms of a classical macroscopic theory on the basis of an effective 
dielectric function, taking account of geometry and anisotropy of samples. 
For instance, the EEL of a multi-wall fullerene can be well explained 
with an effective medium in which the dielectric constant of the 
normal direction to the multi-walls is different from that of the 
other direction.\cite{Lucas}

There are extensive references on the EEL spectrum
 for various geometrical 
objects\cite{Ritchie,Schmeits,Rivacoba,Bertsch}, though most works 
neglect the retardation effect. 
As for the composite  materials, Maxwell-Garnett(MG)-type approximations 
have been widely used.  In the last decade 
a remarkable progress has been made on this subject.   
Pendry et al. proposed a new theoretical method for the EEL of 
periodically arranged nano-structures\cite{PendryMoreno} in terms of 
the real-space 
transfer matrix of the electro-magnetic wave.\cite{PendryMac} 
Garc\'{\i}a de Abajo et al. developed a multiple-scattering method  
for clusters of nano-particles, on the basis of  vector spherical waves
\cite{Abajo1,Abajo2}
and of the boundary element method.\cite{Abajo3,Abajo4}  
These methods are very powerful and in principle can be adapted 
to various complex geometries. However, owing to the spatial discretization 
used in the transfer matrix method and in the boundary element 
method, the accuracy of the results reduces to some extent at high 
frequencies.

Here, we present a fully relativistic multiple-scattering approach  
focused on  arrays of non-overlapping cylinders 
by using vector cylindrical waves. 
Though at present the method works only for the cylinders with isotropic 
dielectric functions, in principle it can work for anisotropic 
cylinders such as multi-wall carbon nano-tubes\cite{Vidal1}. 
With this method the light scattering, as well as the EEL and the induced 
radiation emission in arbitrary arrays of cylinders can be treated 
in a unified manner 
by solving exactly the multiple-scattering equation.  
In addition, since the method utilizes the Fourier decomposition along 
the cylindrical axis, the local optical responses of a certain wave vector 
along the axis can be obtained. This property is feasible in order to 
see the effects of the localized electro-magnetic modes of cylinders 
in detail.

A certain amount of the EEL in composite materials 
is caused by  the induced radiation emission. 
As for a metallic nano-particle, the phenomena is known as the surface plasmon 
radiation emission.\cite{Yamamoto}
 By arranging nano-particles periodically,  
so-called Smith-Purcell(SP) radiation\cite{SP,Abajo5} takes place.  
One of the authors (K.O.) and his collaborators have studied the SP radiation 
in the photonic crystals composed of dielectric spheres in detail, where 
a notable enhancement of the intensity of the SP radiation 
takes place owing to the 
singular state density of photon in the photonic crystals.
\cite{OhtakaYama,Yamaguti1,Yamaguti2} 
However, to the best of our knowledge, none has been reported concerning 
the radiation emission from arrays of cylinders.     
Thus,  the quantitative evaluation of the radiation emission in the arrays   
is another theme of the paper.

On the other hand, the determination of the effective dielectric function 
of composite materials is still valuable at low frequencies. 
It gives a concise explanation of the absorption spectrum as well as 
the EEL spectrum of the composite materials in bulk.\cite{Landau} 
 Moreover, the determination is an important issue  
in the field of left-handed material, which has  
negative permittivity and permeability simultaneously.  
Several methods to  determine them  have been proposed by many authors.
\cite{MG,Pitarke,Smith} 
Here, we propose an alternative method by using the scattering matrix of 
semi-infinite photonic crystals and compare the EEL in the effective medium 
with that of photonic crystal.

This paper deals mainly with the formalism of our method as well as 
the numerical analysis on clusters of Aluminum cylinders with a diameter of 
a few nano-meters, bearing carbon nano-tube arrays in mind. 
In the paper II  we will deal with a metallic 
photonic crystal whose lattice constant is comparable with or larger than 
the plasma wavelength of the constituent metallic cylinders. 
Since it is very bulky to present a comprehensive analysis on the EEL and the 
induced radiation emission in a single paper, 
we should discuss the above topics separately. 
In a nano-structure, which is discussed in this paper, 
the relevant range in wavelength, 
which is near the surface plasma wavelength of metal, is 
much larger than the diameter and than the pitch of the structure.   
As a result,  effects of usual photonic bands, which come from zone 
foldings and degeneracy shifts on Bragg planes,  
do not appear in the EEL spectrum and 
the induced radiation emission spectrum at relevant frequencies. 
In particular there is no remarkable feature in the SP radiation emission 
spectrum. 
Instead, unusual photonic bands of coupled surface plasmon polaritons  
have a strong influence on the EEL spectrum at very low frequencies. 
Moreover, in the relevant frequency range an effective medium approximation 
can be reasonably adapted to the structure. 
As for the  structure which will be discussed in the paper II, 
we will  show that effects 
of both the usual photonic bands 
are very pronounced for the EEL and the induced radiation emission spectra.

The paper is organized as follows. In Sec.II we briefly summarize the vector
cylindrical wave formalism. Using the formalism 
the dispersion relation of the surface plasmon polariton in an isolated 
 metallic cylinder is obtained. 
The multiple scattering method is adapted to the EEL and the induced radiation 
emission in clusters of metallic cylinders in Sec.III. 
In Sec. IV the expression of the EEL in metallic photonic crystals 
is derived. The numerical results of the EEL spectrum 
are compared both with those of 
the isolated cylinder and with those of an effective homogeneous medium. 
Finally, we summarize the results.


\section{Surface plasmon polariton in an isolated cylinder}

An infinitely long metallic cylinder with a circular cross section 
can support an electro-magnetic  surface-localized mode 
that is called surface plasmon polariton(SPP). The SPP is characterized 
by an angular momentum $l$, a wave number $k_z$, and an angular frequency
 $\omega$, 
owing to the rotation invariance with respect 
to the cylindrical axis, the translational invariance along the axis, and 
the translational invariance of time, respectively.  
Before considering its dispersion relation, it is valuable to note some 
formulas of the light scattering by an isolated cylinder using vector 
cylindrical waves.\cite{OhtakaUeta}  
From now on, we take the cylindrical axis to the z-axis. 
Assume that a monotonic incident wave with a momentum $k_z$ along 
the cylindrical axis is scattered by an isolated cylinder with 
a dielectric function $\varepsilon_a(\omega)$ and a radius $r$ embedded 
in a host with a permittivity $\varepsilon_b$. 
Throughout the paper  $\varepsilon_b$ is taken to be 1 in numerical 
calculations, though we keep it unspecified in equations. 
The incident wave can be written as a superposition of vector 
cylindrical waves: 
\begin{eqnarray}
& &{\bf E}^0({\bf x})=e^{ik_zz}\left[ \left(-{1\over\lambda_b}\hat{z}\times\nabla_\|\right)
\psi^{M,0}({\bf x}) \right.\nonumber\\
& &\hskip60pt \left. +\left({ik_z\over \lambda_b q_b}\nabla_\|
+{\lambda_b\over q_b}\hat{z}\right)\psi^{N,0}({\bf x}) \right],\\
& &\psi^{\beta,0}({\bf x})=\sum_l J_l(\lambda_b\rho)e^{il\theta}
\psi_l^{\beta,0} \quad (\beta=M,N),\\
& &\lambda_b=\sqrt{q_b^2-k_z^2},\quad q_b=\sqrt{\varepsilon_b}{\omega\over c},
\end{eqnarray}
where $c$ is the speed of light in vacuum, $(\rho,\theta,z)$ is the 
cylindrical coordinate, $\hat{z}$ is the unit vector along 
the $z$-direction, and 
$\nabla_\|$ is the differential operator with respect to ($\rho,\theta$).  
Throughout the paper, we take the following convention of the square root of 
a complex number:  ${\rm Im}\sqrt{w}\ge 0$ for ${\rm Im}(w)\ge 0$. 
Sometimes, we call the M(N)-field the P(S) polarization.   
At $k_z=0$ the M(N)-field corresponds to the TE(TM) polarization.

A metallic or dielectric cylinder scatters light irrespective of whether 
the light is evanescent or not. By imposing the boundary condition of 
Maxwell's equation, we can solve the scattering problem exactly.  
The induced wave scattered by the cylinder is given by
\begin{eqnarray}
& &{\bf E}^{\rm ind}({\bf x})=e^{ik_zz}\left[ 
\left(-{1\over\lambda_b}\hat{z}\times\nabla\right)
\psi^{M,{\rm ind}}({\bf x}) \right. \nonumber\\
& &\hskip60pt +\left. \left({ik_z\over \lambda_b q_b}\nabla_\|
+{\lambda_b\over q_b}\hat{z}\right)\psi^{N,{\rm ind}}({\bf x}) \right],\\
& &\psi^{\beta,{\rm ind}}({\bf x})=\sum_{l} H_l(\lambda_b\rho)e^{il\theta}
\psi_l^{\beta,{\rm ind}} ,\\
& &\psi_l^{\beta,{\rm ind}} =\sum_{\beta'}t_l^{\beta\beta'}\psi_l^{\beta,0} 
\label{singlescat}
\end{eqnarray}
Here, $H_l$ is the Hankel function of first kind and 
$t_l^{\beta\beta'}$ is the {\it t-matrix} of the cylinder, 
its analytical expression being 
\begin{widetext}
\begin{eqnarray}
& &\left(\begin{array}{cc}
t_l^{MM} & t_l^{MN} \\
t_l^{NM} & t_l^{NN}
\end{array}\right)=-\left(\begin{array}{cc}
d_l^{>MM} & d_l^{>MN} \\
d_l^{>NM} & d_l^{>NN}
\end{array}\right)\left(\begin{array}{cc}
d_l^{<MM} & d_l^{<MN} \\
d_l^{<NM} & d_l^{<NN}
\end{array}\right)^{-1}, \label{tmat}\\
& &\left(\begin{array}{cc}
d_l^{>MM} & d_l^{>MN} \\
d_l^{>NM} & d_l^{>NN}
\end{array}\right)=\left(\begin{array}{cc}
\rho_b J_l'(\rho_a)J_l(\rho_b)-\rho_a J_l(\rho_a)J_l'(\rho_b) & 
l{k_z\over q_a}({\rho_b\over\rho_a}-{\rho_b\over\rho_a}) J_l(\rho_a)J_l(\rho_b)\\
l{k_z\over q_b}({\rho_b\over\rho_a}-{\rho_b\over\rho_a}) J_l(\rho_a)J_l(\rho_b)&{q_a\over q_b}\rho_b J_l'(\rho_a)J_l(\rho_b)-{q_b\over q_a}\rho_a J_l(\rho_a)J_l'(\rho_b)  
\end{array}\right),\\
& &\left(\begin{array}{cc}
d_l^{<MM} & d_l^{<MN} \\
d_l^{<NM} & d_l^{<NN}
\end{array}\right)=\left(\begin{array}{cc}
\rho_b J_l'(\rho_a)H_l(\rho_b)-\rho_a J_l(\rho_a)H_l'(\rho_b) & 
l{k_z\over q_a}({\rho_b\over\rho_a}-{\rho_b\over\rho_a}) J_l(\rho_a)H_l(\rho_b)\\
l{k_z\over q_b}({\rho_b\over\rho_a}-{\rho_b\over\rho_a}) J_l(\rho_a)H_l(\rho_b)&{q_a\over q_b}\rho_b J_l'(\rho_a)H_l(\rho_b)-{q_b\over q_a}\rho_a J_l(\rho_a)H_l'(\rho_b)  
\end{array}\right),\\
& &\rho_i=\lambda_i r \quad(i=a,b),\quad \lambda_a=\sqrt{q_a^2-k_z^2},
\quad q_a=\sqrt{\varepsilon_a}{\omega\over c}
\end{eqnarray}
\end{widetext}
It should be emphasized that except for $k_z=0$ or $l=0$, 
the M(N) polarization 
mixes with the N(M) polarization in the induced wave.  
This property is quite distinct from the case of an isolated sphere, where 
the polarization mixing does not take place. 
In the light scattering by cylinder there is another distinct point from that 
by sphere. Since $\lambda_b$ is the wave number in the $(x,y)$ plane, 
the induced wave 
is evanescent when $\lambda_b$ is pure imaginary. On the other hand 
the induced wave by the sphere behaves as $h_l(q_b|{\bf x}|)$, where $h_l$ is 
the spherical Hankel function of first kind.  This means 
that induced wave by the sphere is always real and has a net flux at 
$|{\bf x}|=\infty$
irrespective of whether the incident wave is evanescent or not.

Next, we focus on the SPP mode in an isolated metallic cylinder.  
The SPP mode is a real eigenmode in the cylinder 
and can exist without the incident light. 
Therefore, taking account of Eq.(\ref{singlescat}) 
the equation that determines 
the dispersion relation of the SPP is given by  
\begin{eqnarray}
\left(\begin{array}{cc}
t_l^{<MM} & t_l^{<MN} \\
t_l^{<NM} & t_l^{<NN}
\end{array}\right)^{-1}
\left(\begin{array}{c}
\psi_l^{M,{\rm ind}}  \\
\psi_l^{N,{\rm ind}} 
\end{array}\right)=0. 
\end{eqnarray}
Like as the SPP on a flat metal/air interface, whose dispersion relation 
 is given by the pole of the interface S-matrix between the metal and air, 
the dispersion relation of the SPP in a metallic cylinder  
is given by the pole of the t-matrix.   
The above equation leads to the following secular equation: 
\begin{eqnarray}
{\rm det}\left(\begin{array}{cc}
d_l^{<MM} & d_l^{<MN} \\
d_l^{<NM} & d_l^{<NN}
\end{array}\right)=0, \label{sppeq}
\end{eqnarray}
for the SPP mode with angular momentum $l$. 
Strictly speaking, we must say that the SPP mode is referred to the mode 
with pure imaginary $\lambda_a$, when the cylinder is of metal. 
The secular equation also determines the dispersion relation 
of the guided modes, in which $\lambda_a$ is real and positive, 
when the cylinder is of dielectric.

For simplicity, from now on, we restrict our consideration to 
an Aluminum cylinder with diameter 2.5[nm], though the formalism presented 
in the paper can be adapted to any types of cylinders as long as an isotropic 
dielectric function of the cylinder is concerned. 
The dielectric function of the Aluminum 
can be  approximated well with the Drude formula: 
\begin{eqnarray}
\varepsilon_a(\omega)=1-{\omega_p^2 \over \omega(\omega+i\eta)}, 
\end{eqnarray}
$\hbar\omega_p$ and $\hbar\eta$ being 15[eV] and 1[eV], respectively. 
We focus on this cylinder embedded in air ($\varepsilon_b=1$) 
in numerical calculations of the paper. 
The small diameter of the cylinder is comparable with that of a carbon 
nano-tube and the plasma frequency is $3.67\times 10^{15}$[Hz] and
 82.7[nm] in terms of wavelength, 
which corresponds to ultra-violet light. Though the cylinder has 
a nano-scale diameter, it seems  still reasonable to 
apply the macroscopic dielectric function to the cylinder.

In Fig.1 the dispersion relation of the SPP modes of the Aluminum cylinder 
is shown.  
\begin{figure}
\includegraphics[scale=0.3]{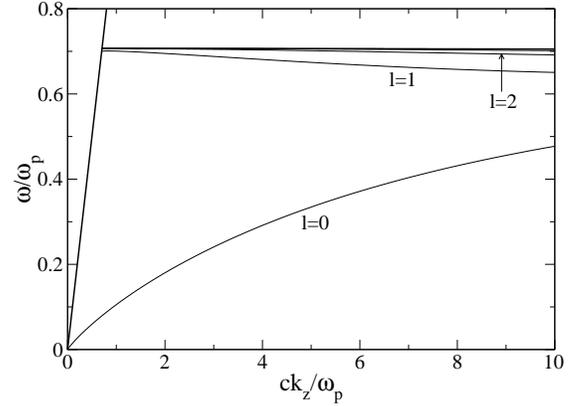}
\caption{\label{}The dispersion relation of the surface plasmon 
polariton of the Aluminum cylinder with diameter 2.5[nm] in air.  
The loss-less Drude dielectric function with $\hbar\omega_p=15$[eV] 
was assumed for the Aluminum cylinder. 
The bold line is the light line.}
\end{figure} 
In this case the SPP mode with $l=0$, which is the solution of 
$d_0^{<NN}=0$, is quite distinct from those with $l\ge 1$. 
The mode with $l=0$ ends at $\omega=0$  and has positive slope in the 
($\omega,k_z$) space, whereas  the other modes 
end at finite $\omega$ on the light line and have negative slopes.   
As $l$ increases, the dispersion curves approach to 
$\omega=\omega_p/\sqrt{2}$. Though the dispersion curves end on the 
light line,   
they can be  extrapolated inside the light-cone.  With this extrapolation  we 
can understand 
how the SPP resonance occurs when real light is incident on the cylinder. 
The resonance also affects significantly the induced radiation emission when 
a charged particle passes near the cylinder.


\section{electron energy loss in cluster of cylinders}

The formalism presented above serves to describe the 
EEL and the induced radiation emission when a charged particle 
runs near a cluster of cylinders.  
In this method the retardation effect is fully taken into account, though 
the recoil of the charged particle is neglected.  In a typical EEL 
experiment the kinetic energy of the electron is about 100 [keV], 
whereas the total energy loss is less than 0.01$\%$ of it, 
so that the assumption of neglecting the recoil  is fairly justified.  
 
Let us consider a charged point particle with a charge $e$ and a velocity $v$ 
runs near the cluster composed of non-overlapping metallic cylinders 
aligned in the $z$-direction,    
whose positions are  specified by two-dimensional ($x$ and $y$) vector 
${\bf x}_\alpha (\alpha=1,2,...,N)$, $N$ being the number of the cylinders.  
For simplicity, we assume that the charged particle 
does not penetrate any cylinders in the cluster and the trajectory of the 
particle is perpendicular to the cylindrical axes. 
Therefore, without losing generality, the position of the particle at time $t$ 
is taken to ${\bf x}_t=(vt,y_0,z_0)$ in the Cartesian coordinate. 
As is known well, a running charged particle is accompanied 
by the electro-magnetic wave that is given by 
\begin{eqnarray}
& &{\bf E}^0({\bf x},\omega)=-{\omega\mu_0 e \over 2} 
\int{dk_z\over 2\pi}{\bf \epsilon}^{\pm}
e^{i{\omega\over v}x+i\gamma|y-y_0|+ik_z(z-z_0)}, \\
& &{\bf \epsilon}^{\pm} = \left(
\begin{array}{c}
{1\over \gamma}\left(1-{1\over q_b^2}({\omega\over v})^2 \right) \\
\mp{1\over q_b^2}{\omega\over v} \\
-{1\over q_b^2}{\omega\over v}{k_z\over \gamma}
\end{array}
\right),\\
& &\gamma=\sqrt{q_b^2-({\omega\over v})^2-k_z^2},
\end{eqnarray}
in the time-Fourier component. 
Here, the superscript of $\epsilon$ is referred to the sign of $y-y_0$. 
If the light velocity in the background medium 
is smaller than $v$, the electro-magnetic wave becomes real, yielding  
the Cerenkov radiation.  Here, we restrict ourselves to the region 
$v<c/\sqrt{\varepsilon_b}$, so that the electro-magnetic wave is evanescent
($\gamma$ is pure imaginary).

The above expression can be transformed 
into a linear combination of the vector cylindrical waves centered at 
${\bf x}={\bf x}_\alpha$ as 
\begin{eqnarray}
& &{\bf E}^0({\bf x},\omega)=\int{dk_z\over 2\pi}e^{ik_z(z-z_0)}
\left[ \left(-{1\over\lambda_b}\hat{z}\times\nabla\right)
\psi_\alpha^{M,0}({\bf x}) \right. \nonumber\\
& &\hskip60pt \left. +\left({ik_z\over \lambda_b q_b}\nabla_\|
+{\lambda_b\over q_b}\hat{z}\right)\psi_\alpha^{N,0}({\bf x})\right],
\label{inc}\\
& &\psi_\alpha^{\beta,0}({\bf x})=\sum_l J_l(\lambda_b|{\bf x}-{\bf x}_\alpha|)
e^{il\theta({\bf x}-{\bf x}_\alpha)}
\psi_{l,\alpha}^{\beta,0},\\
& & \psi_{l,\alpha}^{M0}= {\mu_0 e\omega\over 2}  
e^{i{\omega\over v}x_\alpha \pm i\gamma (y_\alpha-y_0)}
\left(\pm{i\over \lambda_b}\right)i^l e^{-il\theta_{K^\pm}}, \label{M-comp}\\ 
& & \psi_{l,\alpha}^{N0}= {\mu_0 e\omega\over 2} 
e^{i{\omega\over v}x_\alpha \pm i\gamma (y_\alpha-y_0)}
\left({k_z\omega\over q_bv\gamma\lambda_b}\right) i^l e^{-il\theta_{K^\pm}}.   
\label{N-comp}
\end{eqnarray}
Here, $\theta_{K^\pm}$ is the argument of two-dimensional vector 
$K^\pm\equiv(\omega/v,\pm\gamma)$ for a real $\gamma$. 
In the case of a pure imaginary  $\gamma$ we must define  $\theta_{K^\pm}$ as  
\begin{eqnarray}
e^{il\theta_{K^\pm}}=\left({\omega/v\pm i\gamma \over \lambda_b} \right)^l
\end{eqnarray} 
Moreover, 
the $\pm$ in Eq.(\ref{M-comp}) and (\ref{N-comp}) corresponds to 
the sign of $y_0-y_\alpha$ and 
$\psi_{l,\alpha}^{\beta 0}$ is the multi-pole coefficient of 
the incident wave for cylinder $\alpha$. 
We should note that the $k_z$  integration is involved in Eq.(\ref{inc}).
 However, since $k_z$ is a conserved quantity for 
the cluster, each cylindrical wave with fixed $k_z$ is independently 
scattered by the cluster.

The incident evanescent wave is multiply scattered in the cluster 
of the cylinders. By using the multiple-scattering method, the induced 
radiation field is self-consistently determined as follows.\cite{Abajo1}  
If we focus on cylinder $\alpha$, the incident wave 
consists of  the direct term (Eq.(\ref{inc})) plus 
the sum of the induced wave scattered by another cylinder $\alpha'$. 
Therefore, the induced wave from cylinder $\alpha$ is obtained by 
multiplying the t-matrix of cylinder $\alpha$ to the multi-pole components 
of the incident wave re-expanded around ${\bf x}={\bf x}_\alpha$.   
As a result, the self-consistent induced wave is determined as 
\begin{eqnarray}
& &{\bf E}^{\rm ind}({\bf x},\omega)=\int{dk_z\over 2\pi}e^{ik_z(z-z_0)}
\left[ \left(-{1\over\lambda_b}\hat{z}\times\nabla\right)
\psi^{M,{\rm ind}}({\bf x}) \right. \nonumber\\
& &\hskip60pt \left. +\left({ik_z\over \lambda_b q_b}\nabla_\|
+{\lambda_b\over q_b}\hat{z}\right)\psi^{N,{\rm ind}}({\bf x})\right],
\label{ind}\\
& &\psi^{\beta,{\rm ind}}({\bf x})=\sum_{l,\alpha} 
H_l(\lambda_b|{\bf x}-{\bf x}_\alpha|)e^{il\theta({\bf x}-{\bf x}_\alpha)}
\psi_{l,\alpha}^{\beta,{\rm ind}},\\
& &\psi_{l,\alpha}^{\beta,{\rm ind}}=\sum_{\beta'}t_{l,\alpha}^{\beta\beta'}
(\psi_{l,\alpha}^{\beta',0}+
\sum_{l'}\sum_{\alpha'\ne\alpha} G_{l\alpha,l'\alpha'}
\psi_{l',\alpha'}^{\beta',{\rm ind}}),\\
& &G_{l\alpha,l'\alpha'}=H_{l'-l}^{(1)}(\lambda_b\rho_{\alpha\alpha'})
e^{i(l'-l)\theta_{\alpha\alpha'}}. 
\end{eqnarray}
Here,  $G_{l\alpha,l'\alpha'}$ is  the propagator from cylinder $\alpha'$ to 
$\alpha$, and in its expression   
$\rho_{\alpha\alpha'}$ and $\theta_{\alpha\alpha'}$ are the magnitude and 
the argument of ${\bf x}_\alpha- {\bf x}_{\alpha'}$, respectively.

As was mentioned in Sec. II, the induced wave contains propagating components 
with real $\lambda_b$, which have a net flux at $\rho\rightarrow\infty$. 
This implies that a radiation emission takes place when a charged particle 
passes near the cluster. 
As a result, the charged particle losses its energy via the emission.  
In addition, if the cylinders are lossy having positive 
imaginary part in $\varepsilon_a(\omega)$,  a part of the energy  
is absorbed in the cylinders. 
The total energy loss is then the sum of the radiation emission and 
the absorption.  Qualitatively, the loss can be calculated with  
the exerted force 
by the induced field reacting on the particle. 
Thus, its expression is given by  
\begin{eqnarray}
P_{el}(\omega)=-{e\over 2}{\rm Re}\int dt e^{-i\omega t}
{\bf v}\cdot{\bf E}^{\rm ind}({\bf x}_t,\omega)
\end{eqnarray}
per unit angular frequency. 
The integral over $t$ yields 
\begin{eqnarray}
& &P_{el}(\omega)=\int{dk_z\over 2\pi}P_{el}(\omega,k_z),\\
& &P_{el}(\omega,k_z)={1\over 2}\mu_0e^2\omega\sum_{l,\alpha}{\rm Re}\left[
e^{-i{\omega\over v}x_\alpha\pm i\gamma (y_0-y_\alpha)} \right. \nonumber \\
& &\hskip10pt\times\left. (-i)^l e^{il\theta_{K^\pm}} 
\left(\pm{i\over \lambda_b}\psi_{l,\alpha}^{M,{\rm ind}}
-{k_z\omega\over vq_b\lambda_b\gamma}
\psi_{l,\alpha}^{N,{\rm ind}}\right)\right]. 
\end{eqnarray}
Here, $\pm$ in the above equation is referred as the sign of $y_0-y_\alpha$.

On the other hand the net flux of the induced radiation emission 
 is obtained by 
\begin{eqnarray}
& &P_{em}(\omega)=\lim_{\rho\rightarrow\infty}
{1\over 2}\int dzd\theta\rho \\
& &\hskip20pt\times{\rm Re}\left[({\bf E}^{\rm ind}({\bf x},\omega))^*\times
{\bf H}^{\rm ind}({\bf x},\omega)\right]\cdot\hat{\rho}. 
\end{eqnarray}
per unit angular frequency. 
Using the asymptotic form of the induced field (Eq.(\ref{ind})), the net flux 
turns out to be  
\begin{eqnarray}
& &P_{em}(\omega)=\int_{-q_b}^{q_b}{dk_z\over 2\pi}P_{em}(\omega,k_z),\\
& &P_{em}(\omega,k_z)={\mu_0e^2\omega\lambda_b\over 8}\int d\theta 
(|f^M(\theta)|^2+|f^N(\theta)|^2),\label{emclus}\\
& &f^\beta(\theta)=\sqrt{{2\over \pi\lambda_b}} \sum_{l,\alpha}
e^{-i\lambda_b\hat{\rho}{\bf x}_\alpha +il\theta}(-i)^{l+1}
\psi_{l,\alpha}^{\beta,{\rm ind}}. 
\end{eqnarray}
The cutoff of the $k_z$ integral comes from the fact that the scattered 
wave becomes evanescent when $|k_z|>q_b$.  
If there is no absorption in the cylinder,  
the energy loss must be equal to the net flux of the radiation emission, 
that is,  
\begin{eqnarray}
P_{el}(\omega,k_z)=P_{em}(\omega,k_z). 
\end{eqnarray}
This equality serves as a criterion of the correctness and the convergence 
of numerically calculated EEL spectrum.

In the case of an isolated cylinder, the integration over $\theta$ in 
Eq.(\ref{emclus}) 
can be performed analytically, yielding 
\begin{eqnarray}
P_{em}(\omega,k_z)={1\over 2}\mu_0\omega e^2 \sum_{l\beta}
|\sum_{\beta'} t_l^{\beta\beta'}\psi_l^{\beta'0}|^2.
\end{eqnarray}

First of all, let's consider the EEL in the isolated Aluminum cylinder. 
The EEL spectrum of the cylinder 
has a sequence of peaks at the frequencies  of the SPP modes. 
This is because the t-matrix has a SPP pole on the real axis in the 
complex plane of $\omega$ as can be seen in Eq.(\ref{tmat}) and (\ref{sppeq}). 
To see the correspondence of the dispersion relation of the SPP, Fig.2 shows  
the integrand $P_{el}(\omega,k_z)$ of the EEL spectrum, changing 
the $k_z$ value from 0 to 10 in units of $\omega_p/c$. 
Here, the velocity and the impact parameter of the charged particle 
were taken to $0.4c$ and $2r$, respectively. 
\begin{figure}
\includegraphics[scale=0.3]{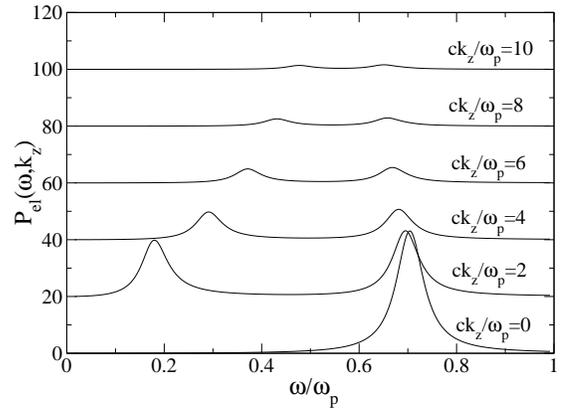}
\caption{\label{}The electron energy loss spectrum in an isolated 
Aluminum cylinder with diameter 2.5[nm] 
for several $k_z$ values are shown. The inverse relaxation time $\eta$ of 
the Drude dielectric function was taken to be $\hbar\eta=1$[eV]. 
The impact parameter, i.e., the distance between the cylindrical axis and 
the trajectory of the charged particle,  is 5[nm]. }
\label{1cylinder}
\end{figure} 
In addition to the main EEL peak around $\omega=\omega_p/\sqrt{2}$, 
another EEL peak appears much below $\omega=\omega_p/\sqrt{2}$. 
This peak comes from the SPP mode with $l=0$, as can be understood 
clearly by comparing with the dispersion relation of the SPP mode 
with $l=0$.      
As a general tendency, $P_{el}(\omega,k_z)$ decreases with increasing  
$|k_z|$.

\begin{figure}
\includegraphics[scale=0.3]{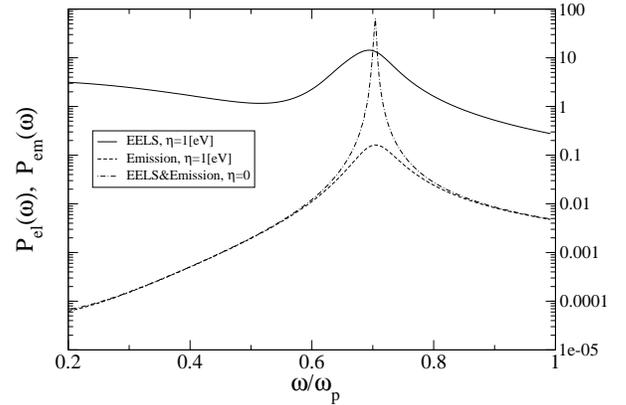}
\caption{\label{}The integrated EEL and the radiation emission spectra 
in the isolated Aluminum cylinder. 
The same parameters as for Fig.\ref{1cylinder} were used.}
\label{1cylinderint}
\end{figure}

Next, we consider how the imaginary part affects the percentage of 
the radiation emission in the EEL. To this end, we show in 
Fig.\ref{1cylinderint} 
the integrated EEL spectrum $P_{el}(\omega)$ and the integrated 
radiation emission spectrum 
 $P_{em}(\omega)$ at $\eta$=0,1[eV]. At $\eta=0$ these must 
coincide. 
As can be seen, when the imaginary part in $\varepsilon_a$ is introduced, 
the radiation emission spectrum is almost unchanged from that in  
the loss-less cylinder  except for 
the SPP frequency region. However, the EEL spectrum receives a drastic 
change.   
According to the figure, the EEL is dominated by the absorption in 
the cylinder under study.     
Therefore,  the efficiency for converting the kinetic energy of 
the charged particle to the radiation  via surface plasmon  
is very low.

\begin{figure}
\includegraphics[scale=0.3]{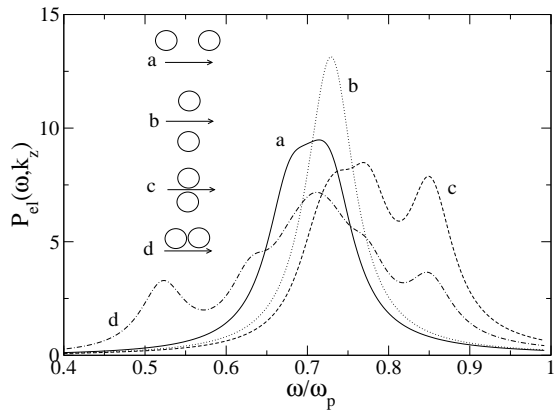}
\caption{\label{}The EEL spectra in the two identical Aluminum 
cylinders with various spatial arrangements shown in the insets. 
$k_z=0$ was assumed. See text for the other parameters.}
\label{2cylinder}
\end{figure}

\begin{figure}
\includegraphics[scale=0.45]{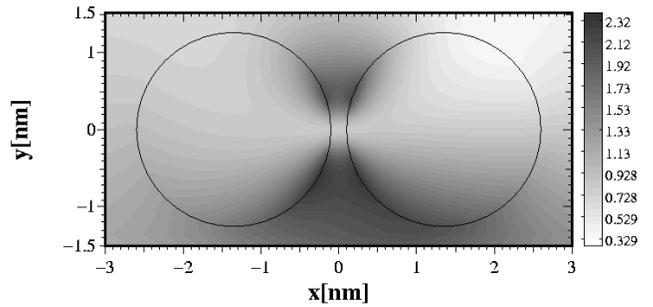}
\caption{\label{}The field intensity $|H_z({\bf x})|^2$ induced by a 
running charged particle whose velocity is $0.4c$ is plotted at 
$(k_z,\omega)=(0, 0.523\omega_p)$, where a peak is observed 
in the EEL spectrum. 
The solid line stands for the edges of the two cylinders. 
The trajectory of the charged particle is at $y=-1.35$[nm].   }
\label{map}
\end{figure}

Next, we consider how the spectra change  when another cylinder is added.  
In this case, the multiple-scattering of the induced radiation significantly 
affects the spectra. As we will see, almost touched metallic cylinders causes 
a drastic change in the  spectra as well as in the near-field 
configuration.  
This phenomena is closely related to the surface-enhanced Raman scattering,  
\cite{Inoue1,Inoue2,Inoue3,Vidal2} in which the intensity of the induced 
electro-magnetic field is enhanced more than thousand times as large as 
that of the incident intensity. 
As an example, we explored the EEL  for 
various spatial arrangements of the two identical Aluminum cylinders. 
Fig.\ref{2cylinder} shows the  EEL spectra  $P_{el}(\omega,k_z)$ with $k_z=0$ 
corresponding to the arrangements shown in the insets. 
In (a) and (b) the two-cylinders are well-separated, the distance 
between two cylindrical axes being $4r$ ($r=1.25$[nm]),   
whereas in (c) and (d) they are very close to each other, the distance being 
$2.16r$.  
The distance between the two cylindrical axes and the trajectory of the charged particle is $2r$ in (a) and (b), and is $1.08r$ in (c) and (d). 
As expected, if the two cylinders are well separated along the trajectory
(case (a)), the EEL spectrum per cylinder has a single peak near 
$\omega=\omega_p/\sqrt{2}$, though an asymmetry of the peak  is observed. 
 In case (b) the two cylinders are separated 
along the normal direction of the trajectory. This geometry  still allows a 
single peak in the EEL spectrum  at $\omega\simeq\omega_p/\sqrt{2}$.  
Since there is the parity symmetry with respect to the trajectory, 
No electro-magnetic modes with the odd parity is involved in case (b). 
On the other hand, if the two cylinders are very close to each other, 
several loss peaks appear    
at the frequencies far from $\omega_p/\sqrt{2}$.
In particular, the spectrum of the case (d) has two marked peaks 
at $\omega/\omega_p\simeq 0.52$ and 0.84. The latter peak is shared also by 
case (c).  The peak at $\omega/\omega_p\simeq 0.52$  is related to the weakly
resonant cavity modes localized in the groove between the two cylinders. 
To confirm this, we show the field intensity  $|H_z({\bf x})|^2$ at 
the peak frequency  in Fig.\ref{map} 
In the figure we can see clearly that the peak is caused by 
the cavity mode in the groove. 
As for the peak at $\omega/\omega_p\simeq 0.84$ 
we couldn't find a clear evidence 
of the resonant cavity mode, as the field intensity at the peak frequency has 
another local maximum at a boundary of the cylinder beside that in the 
grooves. This feature suggests that this peak to be caused by a strong 
mixing of a cavity mode and the SPP modes. 
If the electron runs across the groove of the almost touched cylinders 
(case (c)), 
the  peak at $\omega\simeq 0.52\omega_p$ disappears owing to the 
symmetry mismatch 
and  the peak at $\omega\simeq 0.84\omega_p$ receives an enhancement.

Regarding the radiation emission spectrum in the two cylinders, 
its features are more or less similar to those in Fig.\ref{2cylinder}, 
though the magnitude of $P_{em}(\omega,k_z)$ is much smaller than that of 
 $P_{el}(\omega,k_z)$ at $k_z=0$.


\section{electron running outside photonic crystal}

A photonic crystal that consists of a periodic array of 
metallic cylinders has a rich spectrum in it, including infinite 
SPP bands for the TE polarization\cite{Ito,Ochiai} and a low-frequency 
cutoff for the TM polarization.\cite{Kuzmiak,Pendry3} 
Combining these properties with a running charged 
particle, the system can react as a novel light emitter. 
In fact, when a charged particle passes near the photonic crystal, 
it induces the emission of real photon as was first pointed out by 
Smith and Purcell for a metallic grating.\cite{SP} 
In the photonic crystal, this phenomena can be interpreted in two ways. 
One interpretation is as follows. 
The incident evanescent wave from the charged particle acquires an Umklapp 
momentum transfer in the photonic crystal, 
thereby coming into the light cone in the $(\omega,k_\|)$
space, $k_\|$ being the wave vector parallel to the boundary of the photonic 
crystal.  As a result, real photon is emitted from the photonic crystal. 
The other interpretation is to regard the phenomena as a coherent 
radiation emission from different cylinders.  
Though these two interpretations are equivalent, the two points of view 
give us a deep insight of the Smith-Purcell(SP) radiation in the 
photonic crystal.

\begin{figure}
\includegraphics[scale=0.3]{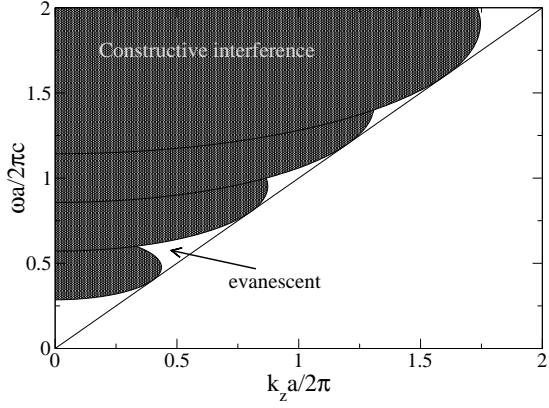}
\caption{\label{} The phase diagram of the induced radiation emission in  
a periodic array of cylinders with lattice constant $a$ is shown. 
The Smith-Purcell radiation is possible 
only in the shaded region of the ($k_z,\omega$) plane.    
The velocity of the charged particle was taken to be
 $v=0.4c$. The solid line is the light line. }
\label{phase}
\end{figure}

In the case of an isolated cylinder, the induced radiation emission 
is possible 
when $q_b^2 > k_z^2$, i.e.,when the emitted light is  inside the light cone 
of the $(\omega,k_z)$ space. 
On the other hand, the condition that the evanescent light turns out to be  
a real photon via the Umklapp momentum transfer in the photonic crystal is 
given by 
\begin{eqnarray}
q_b^2-k_z^2-({\omega\over v}-n{2\pi\over a})^2 >0, 
\end{eqnarray}
$a$ being the pitch of the photonic crystal. 
Therefore, the allowed frequency range of the SP radiation is
\begin{eqnarray}
& &\tilde{\omega}_- <\tilde{\omega} <\tilde{\omega}_+, 
\quad \tilde{k_z}^2 < {n^2\varepsilon_b \over ({c\over v})^2-\varepsilon_b },\\
& &\tilde{\omega}_\pm={n{c\over v}\pm\sqrt{n^2\varepsilon_b-
( ({c\over v})^2-\varepsilon_b)
\tilde{k_z}^2}  \over ({c\over v})^2-\varepsilon_b}, \label{cutoff}
\end{eqnarray}
where $\tilde{\omega}=\omega a/2\pi c$ and  $\tilde{k_z}=k_z a/2\pi$.   
If the above condition is not satisfied in the light cone, 
a destructive interference among the induced radiation fields
 from different cylinders 
occurs, leading to the prohibition of the radiation emission. 
This also implies that if there is no absorption in the photonic crystal, 
the concerned range in the ($\omega,k_z$) space does not contribute 
to the EEL in the photonic crystal. 
The phase diagram of the radiation emission at $v=0.4c$ is shown in 
Fig.\ref{phase}. 
In Fig.\ref{phase} the SP radiation is allowed in the shaded regions. 
The allowed region becomes narrow in the ($\omega,k_z$) space as 
the velocity of the particle decreases. 
Beside, in the allowed region of the SP radiation 
the photonic density of state is in general  
singular including the Van Hove  singularity.  
Therefore, we may expect a quite rich spectrum of the SP radiation 
in the photonic crystals.

The rich spectrum is not limited in the SP radiation. 
We should mention that the photonic band structure exists also outside the 
shaded region of Fig.\ref{phase}.  
As long as the imaginary part in $\epsilon_r$ is non-zero, the absorption of 
the induced radiation is inevitable in the photonic crystal. 
This causes the EEL outside the shaded region, and the EEL is affected 
 significantly by the photonic band structure therein.  
Inside the shaded region the EEL consists of the radiation emission and the 
absorption, and these are independent physical observables.

Like as the case of cluster of  cylinders, the EEL and the SP radiation 
in the photonic crystals 
can be treated in a unified framework with the multiple-scattering 
method on the basis of vector cylindrical waves.  
Pendry and Martin-Moreno presented for the first time 
an unified framework in terms of the transfer matrix method 
to argue the EEL  and the SP radiation in photonic crystals.\cite{PendryMoreno}, 
The method is based on a discretization of Maxwell's equation on a 
spatial mesh in order to obtain the real-space transfer matrix.\cite{PendryMac} 
 They found that the scattering matrix, which is obtained from 
the transfer matrix,  
is directly related to the EEL and SP radiation spectra.   
Their algorithm is easily adapted to the two-dimensional counterpart
\cite{OhtakaNumata,OhtakaUeta,Botten} of   
the layer Korringa-Kohn-Rostoker-Ohtaka (KKRO) method
\cite{Korringa,Kohn,Ohtaka1,Ohtaka2,Modinos,Stefanou}, 
which is a generalization of the 
multiple scattering method to periodic systems. 
The layer-KKRO method has very high accuracy  
for the photonic crystal under consideration.  
We should note that the three-dimensional layer-KKRO method  was already 
used for discussing the EEL and the SP radiation 
in three-dimensional photonic crystals composed of spheres.
\cite{OhtakaYama,Yamaguti1,Yamaguti2,Abajo5}

\begin{figure}
\includegraphics[scale=0.3]{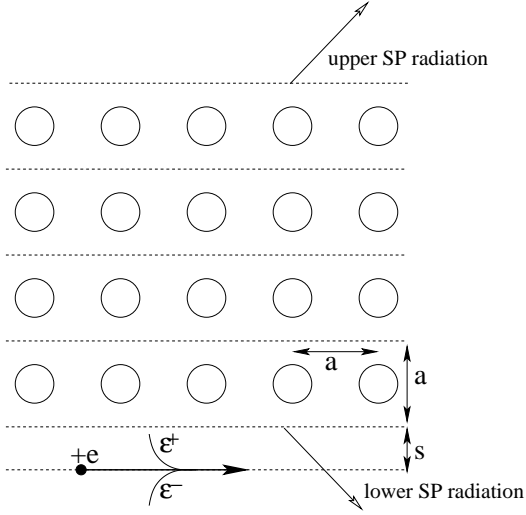}
\caption{\label{} A charged particle runs with distance $s$ from the 
boundary of the finite-thick photonic crystal, which is composed of a square 
array of cylinders with pitch $a$. The polarization vector of the evanescent
 wave accompanied by the charged particle is denoted by $\epsilon^\pm$. }
\label{outside}
\end{figure}

Let's assume that a running charged particle passes outside a finite-thick 
photonic crystal composed of a periodic array of the Aluminum cylinders. 
The particle runs with distance $s$ from the boundary normal to the 
$\Gamma-X$ direction.  
The schematic illustration of the system under study is shown in 
Fig.\ref{outside}. 
In this case the induced radiation field in the outer region of 
the photonic crystal including the particle trajectory is given by 
\begin{eqnarray}
& &{\bf E}^{\rm ind}({\bf x},\omega)=\sum_{h} \int {dk_z\over 2\pi}
e^{i{\bf K}_h^-\cdot{\bf x}}Q_{-+}(h,h_0){\bf \epsilon}^+, \\
& &{\bf K}_h^\pm=(k_x+h,\pm\gamma_h,k_z),\\
& &{\omega \over v}=k_x+h_0,\\ 
& &\gamma_h=\sqrt{q_b^2-(k_x+h)^2-k_z^2}, 
\end{eqnarray}
where $Q_{\pm\pm}(h,h')$, which has spatial tensor index, 
is the scattering matrix of the photonic crystal,\cite{Ohtaka1} 
$k_x$ is the Bloch momentum in the irreducible surface Brillouin zone 
associated with the boundary  
and $h(=2\pi {\bf Z}/a)$ is a reciprocal lattice vector.  
In the opposite outer region of the photonic crystal the induced radiation 
field (upper SP radiation in Fig.\ref{outside}) is given by 
\begin{eqnarray}
& &{\bf E}^{\rm ind}({\bf x},\omega)=\sum_{h} \int {dk_z\over 2\pi}
e^{i{\bf K}_h^+\cdot{\bf x}}Q_{++}(h,h_0){\bf \epsilon}^+.
\end{eqnarray}
Quantitatively, the EEL  per unit length of the particle trajectory 
is expressed by the following equation:
\begin{eqnarray}
& &P_{el}(\omega,k_z)={1\over 4}{\mu_0 e^2\omega} e^{-2|\gamma_{h_0}|s}
\nonumber \\ 
& &\hskip10pt\times|\gamma_{h_0}|{\rm Im}\left[({\bf \epsilon}^+)^\dagger 
Q_{-+}(h_0,h_0){\bf \epsilon}^+\right].
\end{eqnarray}
On the other hand the SP radiation spectrum per unit length is given by 
\begin{eqnarray}
& &P_{em}(\omega,k_z)={1\over 8}{\mu_0 e^2\omega}e^{-2|\gamma_{h_0}|s}
\nonumber \\
& &\times \sum_{h\in {\rm open}}\gamma_h 
( |Q_{++}(h,h_0){\bf \epsilon}^+|^2 + |Q_{-+}(h,h_0){\bf \epsilon}^+|^2 ),
\end{eqnarray}
where the summation is taken over the open diffraction channels.

If the cylinders are loss-less in the photonic 
crystal, 
again $P_{el}(\omega,k_z)=P_{em}(\omega,k_z)$. 
This can easily be confirmed by considering flux conservation of 
through the photonic crystal:
\begin{eqnarray}
& &\sum_{h\in{\rm open}}\gamma_h \left( 
|Q_{++}(h,h_0){\bf \epsilon}^+|^2 + |Q_{-+}(h,h_0){\bf \epsilon}^+|^2 \right)
 \nonumber \\
& & \hskip50pt =2 |\gamma_{h_0}|{\rm Im}\left[({\bf \epsilon}^+)^\dagger Q_{-+}(h_0,h_0){\bf \epsilon}^+ \right]
\end{eqnarray}

\begin{figure}
\includegraphics[scale=0.3]{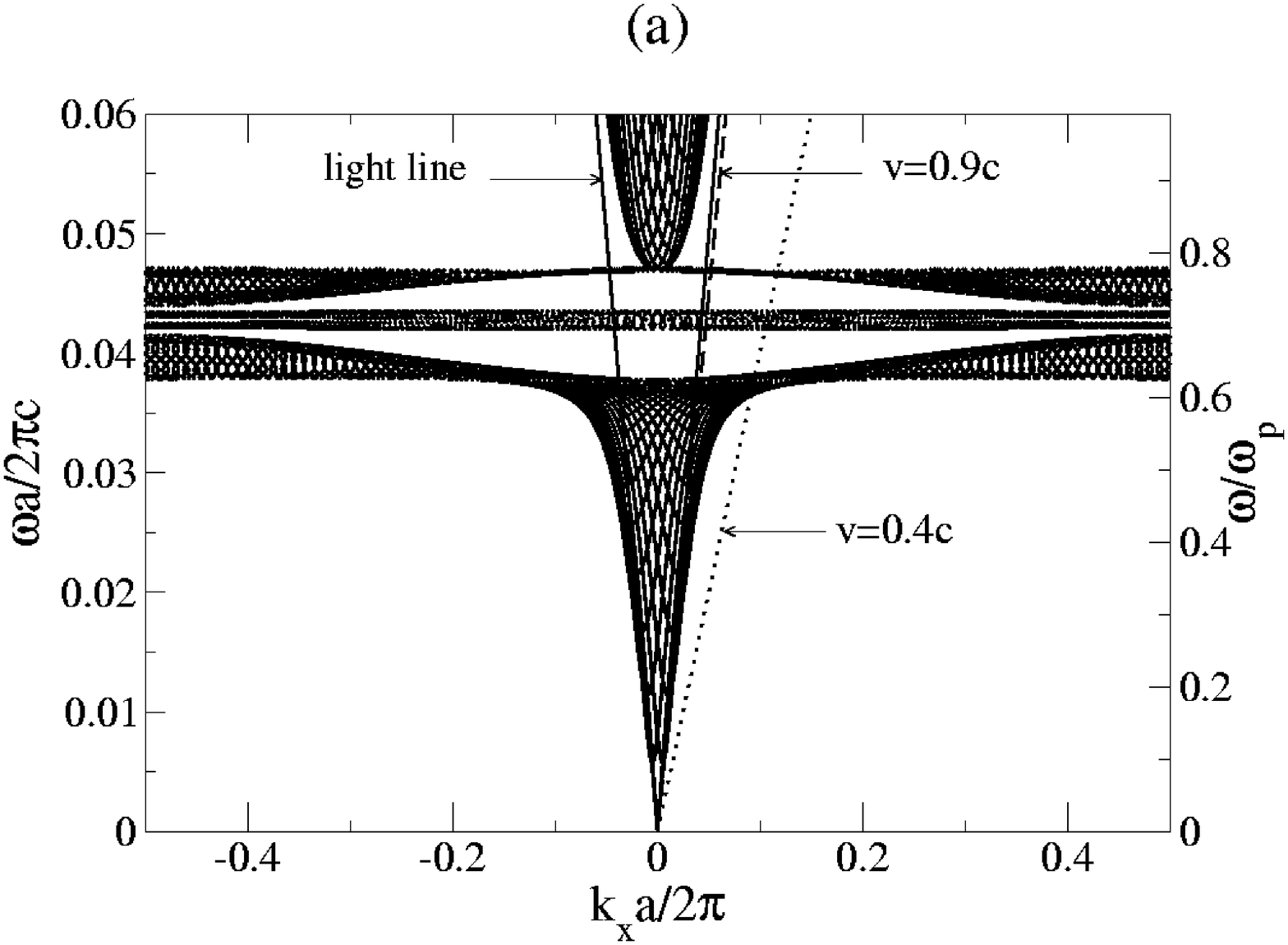}
\vskip30pt
\includegraphics[scale=0.3]{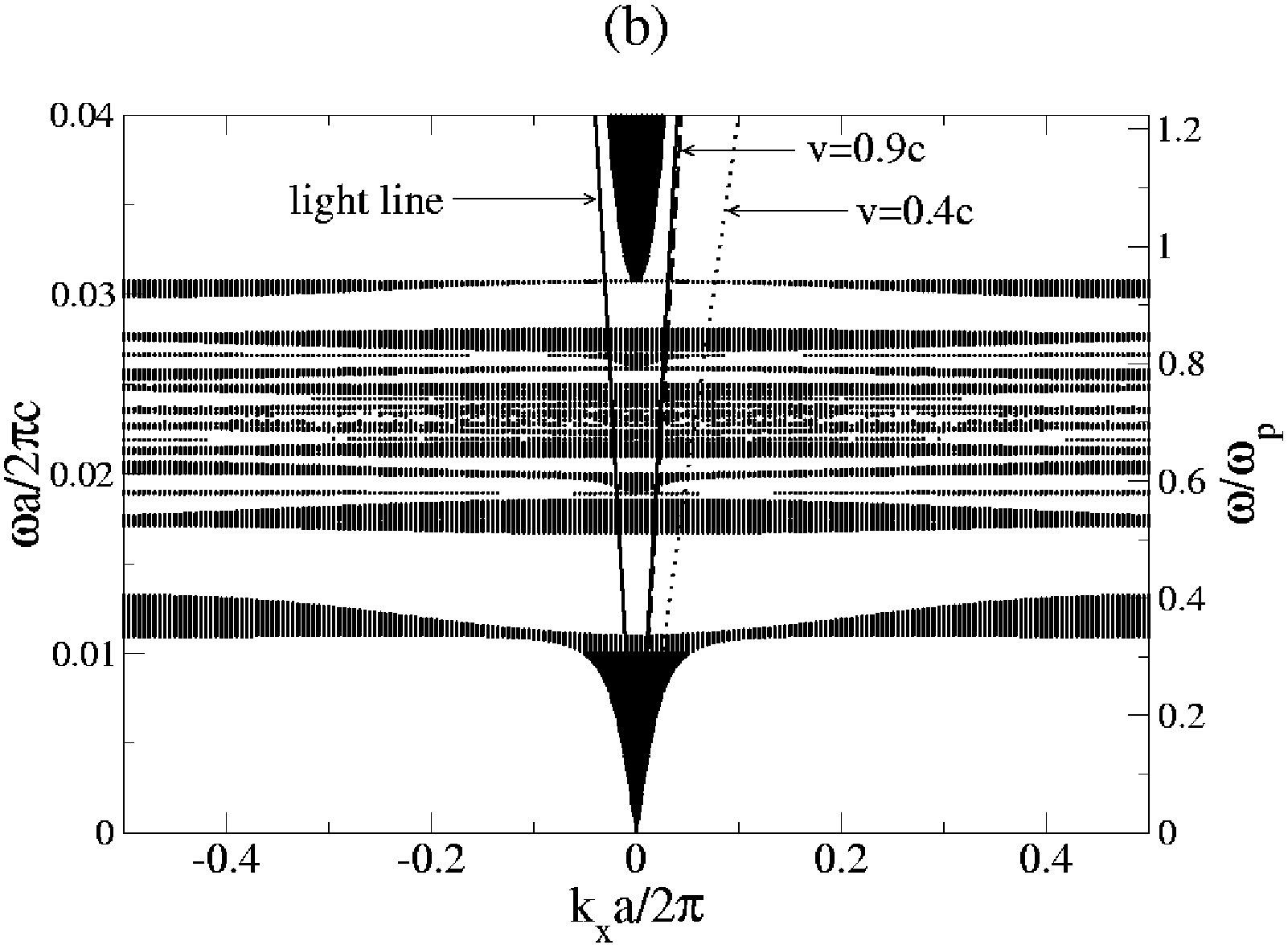}
\caption{\label{}The photonic band structure of the square lattice 
of the Aluminum cylinders at $k_z=0$ was projected on the surface 
Brillouin zone 
associated with the boundary normal to the $\Gamma-X$ direction. 
In (a) the cylinders 
are well separated(lattice constant $a=4r$=5[nm]), whereas in (b) 
the cylinders nearly touch($a=2.16r$=2.7[nm]).}
\label{pbd}
\end{figure}

Here, we study two kinds of metallic photonic crystals. 
One is a dilute photonic crystal that is composed of the  
square array of the Aluminum cylinders with the lattice constant $a=4r=5$[nm]. 
The other is its dense version which corresponds to the cases (c) and (d) of the coupled two cylinders treated  
in the last section ($a=2.16r=2.7$[nm]).  
The photonic band structures of these photonic crystals at $k_z=0$ 
projected on the 
surface Brillouin zone associated with the boundary normal to the 
$\Gamma-X$ direction are shown in Fig.\ref{pbd}.
The band structures were calculated by using the two-dimensional layer-KKRO 
method taking $l_{max}=5$ and 12 in the dilute and dense photonic crystals, 
respectively. 
Here, we dropped the band diagram of the S(TM)-polarization, because 
it is not relevant to our problem. However, at non-zero $k_z$ we must take 
account of both the S and P polarizations. 

In both the photonic crystals the plasma frequency of the cylinder 
is much smaller than the lattice scale, so that 
the photonic band structure is very close to that of the empty lattice at high 
frequencies. However, below $\omega=\omega_p$  many flat bands 
which characterize the  metallic photonic crystals appear.  
These bands are generally anisotropic, reflecting the $C_{4v}$ symmetry of 
the square lattice, and have a singular state density.
In the dilute photonic crystal, many bands are concentrated near 
$\omega=\omega_p/\sqrt{2}$. This indicates that they are merely a 
tight-binding coupling of the SPP of the isolated cylinder. In principle, 
we can find infinite numbers of the flat bands at  $\omega=\omega_p/\sqrt{2}$, 
which are quite difficult to distinguish. 
On the other hand, in the dense photonic crystal 
 the flat bands are diverse in frequency, 
whereas their center is still at $\omega=\omega_p/\sqrt{2}$. 
Some of the bands are originated from  the cavity mode localized in the 
groove of the two cylinders. However, most flat bands are considered to 
be related with the SPP of an Aluminum cylinder. 
 
In Fig.\ref{pbd} the dispersion lines of the charged particle 
($k_x=\omega/v +h_0$) at  
$v=0.4c$ and $0.9c$ as well as the light line ($\omega=ck_x$) 
were also plotted. 
In the frequency range concerned only the line with $h_0=0$ is relevant, 
because the threshold of the SP radiation occurring along the Umklapp shifted 
line of $h_0=2\pi/a$  is rather high 
($\tilde{\omega}_-\simeq 0.286$ and $0.474$ in Eq.(\ref{cutoff}) 
for $v=0.4c$ and $0.9c$, respectively). 
As a result, the EEL is caused solely by the absorption 
in the frequency range concerned. 
When the dispersion line meets the shaded region of the projected band 
diagram, the charged particle can excite an eigenmode in the photonic 
crystal and thus causes an enhanced absorption loss in EEL. 
Strictly speaking, the projected band scheme should be used to understand 
the feature  of the photonic crystal  with infinite 
thickness along the $\Gamma-X$ direction. 
Since we are considering a finite-thick 
photonic crystal, the shaded region in Fig.\ref{pbd} must be 
regarded as a set of the 
dispersion curves of the eigenmodes in the finite-thick photonic crystal. 
Apparently, 
as the thickness increases, the dispersion curves fills up with 
the shaded region.

\begin{figure}
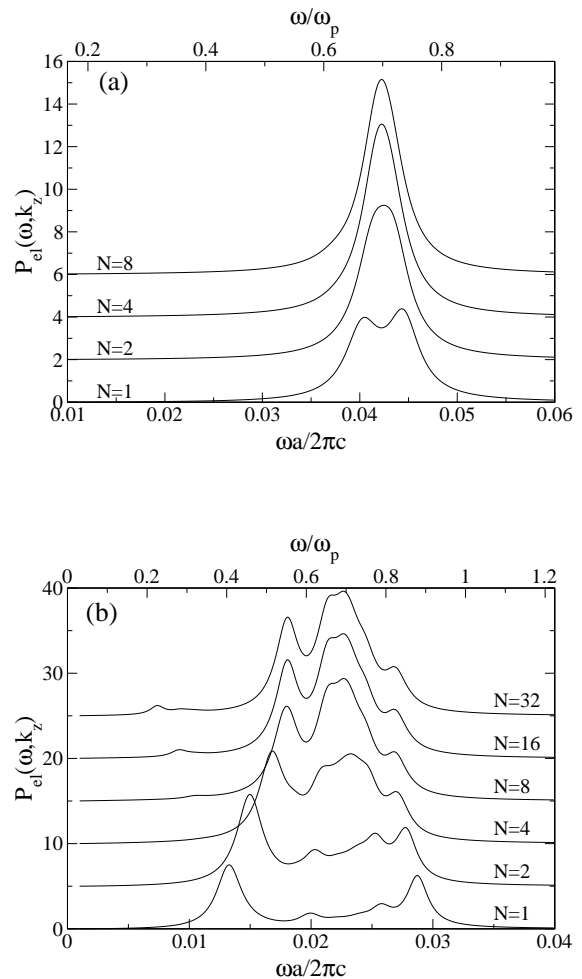

\includegraphics[scale=0.3]{fig9a.eps}
\vskip30pt
\includegraphics[scale=0.3]{fig9b.eps}
\caption{\label{}The EEL spectrum of the photonic crystals at $k_z=0$, 
varying the number of layers. 
The velocity of the charged particle was taken to $v=0.4c$.
In (a)  the cylinders are well separated(lattice constant $a=4r$), 
whereas in (b) the cylinders nearly touch($a=2.16r$). 
The trajectory of the charged particle is just on the boundary of 
the photonic crystal (i.e. $s=0$).}
\label{v=0.4}
\end{figure}

Fig.\ref{v=0.4} shows $P_{el}(\omega,k_z)$ with $k_z=0$ 
of the two photonic crystals,  varying number of layers.
The velocity of the charged particle was taken to $0.4c$ and the 
impact parameter $s$ was taken to zero.    
Concerning the dilute photonic crystal, the EEL spectrum has the double peaks 
near $\omega=\omega_p/\sqrt{2}$ in the mono-layer case. 
This feature already appeared in the result of the two separated cylinders 
(see Fig.\ref{2cylinder}), where an asymmetry of the loss peak is observed. 
As the number of 
layers increases, the double peaks disappear and the spectrum converges 
to a certain function which has single peak near $\omega=\omega_p/\sqrt{2}$. 
The converged spectrum is not so far from the EEL spectrum of the isolated 
cylinder.  These features are consistent with the numerical results on the 
projected band structure (Fig.\ref{pbd}): The dispersion line of the particle 
with $v=0.4c$ meets only the flat bands near 
$\omega=\omega_p/\sqrt{2}$. We can infer that the single peak in the 
EEL spectrum is caused by 
the broadening of the flat bands owing to the non-zero imaginary part 
in $\varepsilon_a$.

As for the dense photonic crystal, 
there are several loss peaks whose positions 
change 
as the number of layers increases. Compared with the case of coupled two 
cylinders, 
the peak positions of the EEL spectrum in the photonic crystal 
are well correlated with  
those of case (d) of Fig.\ref{2cylinder}). 
In particular, the two peaks at $\omega\simeq 0.55\omega_p$ 
and $0.82\omega_p$ at $N=32$ are of reminiscences of those in case (d), 
and the corresponding flat bands, which have relatively large widths in 
frequency,  can be observed in Fig.\ref{pbd}(b).  
Again, above $\omega a/2\pi c\simeq 0.01$ the EEL spectrum converges 
to a certain function with increasing number of layers, 
though the convergence progresses slowly compared with that 
in the dilute photonic crystal. 
A remarkable feature in this case appears below $\omega a/2\pi c=0.01$, where 
 a frequency  shift of a small loss peak is observed with increasing $N$.   
In contrast to the dilute photonic crystal, 
in such low frequency region the dispersion line $\omega=vk_x$ exists within 
the shaded region of the lowest band even at $v=0.4c$. 
This band does not originate from the SPP modes, and thus 
the loss peaks found below $\omega a/2\pi c$  is different 
in feature from that by 
the SPP bands.

The effects of the lowest band in the EEL spectrum can be clearly demonstrated 
in the dilute photonic crystal with large thickness, using impinging 
a charged particle impinging with such high-speed that its
dispersion line $\omega=vk_x$ is in the lowest band in Fig.\ref{pbd}
(a). 
\begin{figure}
\includegraphics[scale=0.3]{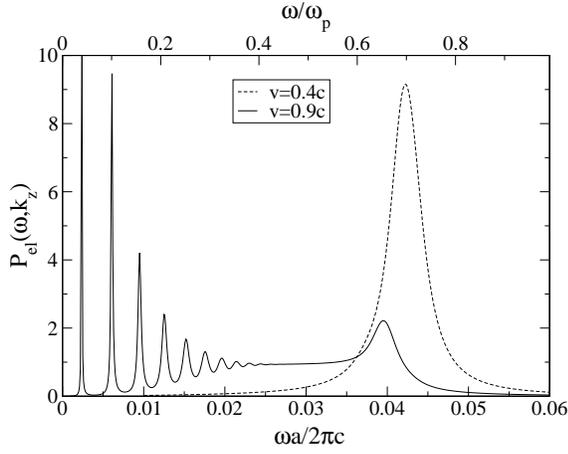}
\caption{\label{}The EEL spectrum of the dilute photonic crystal having 
 256 layers at $k_z=0$. The velocity of the charged particle was 
taken to $0.4c$ (dashed line) and $0.9c$ (solid line). }
\label{v=0.9}
\end{figure} 
Fig.\ref{v=0.9} shows the two EEL spectra of $v=0.4c$ and $0.9c$ 
in the dilute photonic crystal having 256 layers. 
As can be seen in the figure, 
the lowest band causes very sharp loss peaks whose positions are distributed 
below $\omega a/2\pi c\simeq 0.025$.  In order to understand this feature, 
we must remark that outside the light cone the band becomes 
the set of the guided modes in the corresponding finite-thick photonic 
crystal. 
Moreover, as mentioned later, we can reasonably introduce 
an effective dielectric 
function $\varepsilon_{\rm eff}(\omega)$, which is very close to 
that of Maxwell-Garnett,\cite{MG}  to the photonic crystal under consideration.
The effective dielectric function of Maxwell-Garnett is given by 
\begin{eqnarray}
& &\varepsilon_{\rm eff}^{\rm MG}(\omega)=\varepsilon_b
\left( 1+{2f\alpha \over 1-f\alpha} \right),\label{MGeff} \\
& &\alpha={\varepsilon_a-\varepsilon_b \over \varepsilon_a+\varepsilon_b},
\end{eqnarray}
$f$ being the filling ration of the cylinders. 
Using this effective dielectric function, 
the dispersion relation of the guided modes in the (loss-less) 
effective medium is determined by 
\begin{eqnarray}
& &1-\left( {-\gamma/\varepsilon_b+\gamma'/{\rm Re}
                    (\varepsilon_{\rm eff}^{\rm MG})
 \over \gamma/\varepsilon_b+\gamma'/{\rm Re}
             (\varepsilon_{\rm eff}^{\rm MG})} \right)^2
\exp(2i\gamma d)=0, \\
& &\gamma =\sqrt{({\omega\over c})^2\varepsilon_b -k_x^2}, \\
& &\gamma'=\sqrt{({\omega\over c})^2{\rm Re}(\varepsilon_{\rm eff}^{\rm MG})
                  -k_x^2}
\end{eqnarray}
$d$ being the thickness of the photonic crystal. 
By imposing the matching condition of frequency $\omega$ and wave vector 
$k_x(=\omega/v)$, the above equation has a sequence of solutions, which  
agree with the positions of the sharp loss peaks of $v=0.9c$ 
in Fig.\ref{v=0.9} fairly well.

\begin{figure}
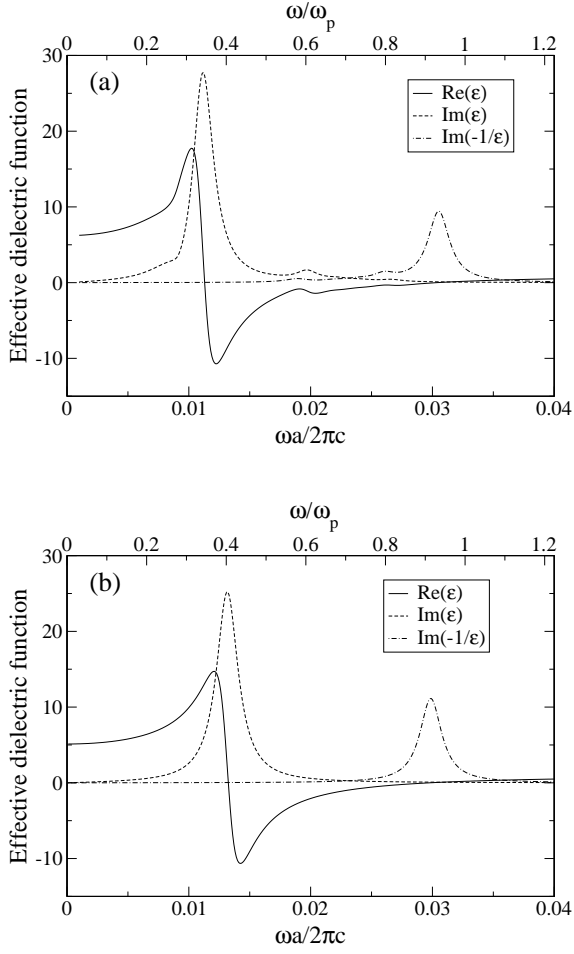

\includegraphics[scale=0.3]{fig11a.eps}
\vskip20pt
\includegraphics[scale=0.3]{fig11b.eps}
\caption{\label{}(a) The effective dielectric function for the P-polarized 
light in the dense photonic crystal($a=2.16r$). The case $k_z=0$ was assumed. 
(b) The effective dielectric function of Maxwell-Garnett in the dense 
photonic crystal.}
\label{effeps}
\end{figure}

The convergence of $P_{el}(\omega,k_z)$ is a direct consequence 
of the convergence of the scattering matrix $Q_{-+}$ itself.  
As was discussed by Botten et al, the converged value of $Q_{-+}$ 
gives the reflectance of the semi-infinite photonic crystal.\cite{Botten}
This also implies that using the converged value of $Q_{-+}$, we can extract 
the effective dielectric function via Fresnel's formula of the interface 
 S-matrix. That is, for the P-polarized incident wave, the scattering matrix 
 $Q_{-+}$ of the semi-infinite photonic crystal can be regarded as the interface S-matrix between the background medium and the effective medium:  
\begin{eqnarray}
& &[Q_{-+}(h_0,h_0)]_{pp}\simeq 
{\gamma/\varepsilon_b-\gamma'/\varepsilon_{\rm eff}\over 
 \gamma/\varepsilon_b+\gamma'/\varepsilon_{\rm eff}}, \label{oureff1}\\
& &\gamma'=\sqrt{({\omega\over c})^2\varepsilon_{\rm eff}
-({\omega\over v})^2}.\label{oureff2}
\end{eqnarray}
Here, $k_z=0$ was assumed. 
The effective dielectric function $\varepsilon_{\rm eff}$ obtained in this 
way, along with that of Maxwell-Garnett 
for the dense photonic crystal are shown in Fig.\ref{effeps}. 
The function is not so far from the effective dielectric function of 
Maxwell-Garnett, though some extra features at $\omega a/2\pi c\simeq 0.019$ 
and $0.026$ are observed. In the next section we will see that 
the EEL spectrum in the photonic crystal, when the charged particle runs 
inside the photonic crystal,  is well  reproduced with the effective 
dielectric 
function having the extra features.  
As for the dilute photonic crystal, our effective dielectric function is 
very close to that of Maxwell-Garnett.


\section{electron running inside photonic crystal}

\begin{figure}
\includegraphics[scale=0.3]{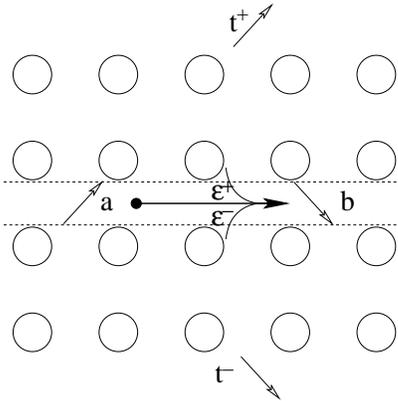}
\caption{\label{} A charged particle runs inside the photonic crystal with an 
equal distance from the upper and lower nearest layers of the trajectory. 
The plane wave coefficients of the induced wave in the void stripe including the trajectory are denoted by {\bf a} and {\bf b}. }
\label{inside}
\end{figure}

When a charged particle runs inside the photonic crystals, 
the induced radiation field is rather involved owing to the 
multiple-scattering among the layers above and below the trajectory.    
However, the scattering matrix formalism is readily adapted to 
the case as long as the particle does not penetrate any cylinders in the 
photonic crystals.   
A schematic illustration of the system under consideration is shown in 
Fig.\ref{inside}.
In this case the induced radiation field reacting back to the charged particle 
is determined as 
\begin{eqnarray}
& & {\bf E}^{\rm ind}({\bf x},\omega)=
\sum_h ({\bf a}_h e^{i{\bf K}_h^+\cdot{\bf x}}+
{\bf b}_h e^{i{\bf K}_h^-\cdot{\bf x}} ),\\
& &{\bf a}_h=(1-Q_{+-}^d Q_{-+}^u)^{-1}
 Q_{+-}^d ({\bf \epsilon}^- +Q_{-+}^u{\bf \epsilon}^+),\\
& &{\bf b}_h=(1-Q_{-+}^u Q_{+-}^d)^{-1}
 Q_{-+}^u ({\bf \epsilon}^+ +Q_{+-}^d{\bf \epsilon}^-),
\end{eqnarray}
where $Q_{\pm\pm}^{u(d)}$ is the scattering matrix of the upper(lower) 
layers above(below) the trajectory.  
Beside, the Fourier coefficients of the upper and lower transmitted wave, 
denoted by ${\bf t}_h^\pm$, is also obtained as
\begin{eqnarray}
& &{\bf t}_h^+=\sum_{h'} Q_{++}^u(h,h') 
({\bf \epsilon}^+\delta_{h'h_0} + {\bf a}_{h'}),\\
& &{\bf t}_h^-=\sum_{h'} Q_{--}^d(h,h') 
({\bf \epsilon}^-\delta_{h'h_0} + {\bf b}_{h'}). 
\end{eqnarray}
Therefore, the EEL and SP radiation spectra per unit length becomes  
\begin{eqnarray}
& &P_{el}(\omega,k_z)={1\over 4}\mu_0 e^2 \omega  
|\gamma_{h_0}|{\rm Im}\left({\bf a}_{h_0}^*\cdot{\bf \epsilon}^- 
-{\bf b}_{h_0}\cdot({\bf \epsilon}^+)^* \right), \\
& &P_{sp}(\omega,k_z)={1\over 8}\mu_0 e^2 \omega 
\sum_{h\in{\rm open}}\gamma_h \left( |{\bf t}_h^+|^2 + |{\bf t}_h^-|^2 \right). \end{eqnarray}
Again, the flux conservation in a loss-less photonic crystal leads 
\begin{eqnarray}
& & \sum_{h\in{\rm open}}\gamma_h 
\left( |{\bf t}_h^+|^2 + |{\bf t}_h^-|^2 \right)
\nonumber \\
& &=-2 |\gamma_{h_0}|{\rm Im}\left({\bf a}_{h_0}^*\cdot{\bf \epsilon}^- 
-{\bf b}_{h_0}\cdot({\bf \epsilon}^+)^* \right), 
\end{eqnarray}
which implies that the EEL is equal to the SP radiation emission.

At low frequencies we may expect that the EEL in a photonic crystal 
is somehow approximated by that of a lossy effective homogeneous medium.   
As is known well, the relativistic EEL in such a  medium with 
permittivity $\varepsilon_{\rm eff}$ is given by 
\begin{eqnarray}
P_{el}(\omega,k_z)={1\over 4}\mu_0 e^2 \omega
{\rm Re}\left[ {1\over \gamma} \left(1-({c\over v})^2
{1\over\varepsilon_{\rm eff}}
\right) \right],\label{lossbulk}
\end{eqnarray}
per unit length. When $\varepsilon_{\rm eff}$ is real and the condition 
$v>c/\sqrt{\varepsilon}$ is satisfied, 
the above equation is equal to the Cerenkov loss.  
Otherwise, Eq.(\ref{lossbulk}) can be regarded as the EEL by the absorption. 
In a homogeneous metal the bulk plasmon dominantly contributes 
to the EEL, because of the factor $1/\varepsilon_{\rm eff}$.

\begin{figure}
\includegraphics[scale=0.3]{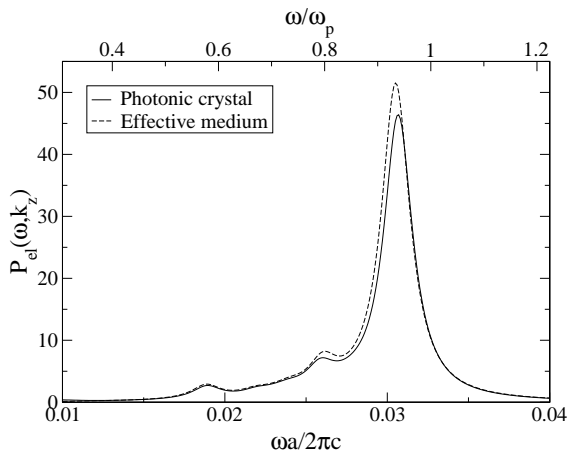}
\caption{\label{} The EEL spectrum in the dense photonic crystal having 64
 layers along the $\Gamma-X$ direction is shown. The EEL spectrum in the 
effective medium whose dielectric function is given by Fig.\ref{effeps}
is also plotted. The velocity of the charged particle is taken to $0.4c$.} 
\label{bulk}
\end{figure}

Fig.\ref{bulk} shows the EEL spectra of the dense photonic crystal and 
its simulation using the effective dielectric function obtained by 
Eqs (\ref{oureff1}) and (\ref{oureff2}). 
Here, the charged particle runs with velocity $0.4c$ 
between the 32th and 33th layer of the 64-layer thick slab of the 
dense photonic crystal.

One can observe that 
the frequency of the main loss peak is larger than $\omega=0.9\omega_p$, 
which is close to the bulk plasmon frequency.
This is quite reasonable, taking account that the dense photonic crystal has 
a large filling ratio ($\simeq 67\%$) and thus is close to the bulk metal of 
Aluminum. Beside the main loss peak, two small peaks are observed in the EEL 
spectrum.  
Using the homogeneous medium approximation with the 
effective dielectric function  obtained in the previous section  
the above features are well reproduced. 
It should be emphasized that  the small two peaks in the dense photonic 
crystal  can not be explained 
with the effective medium theory of Maxwell-Garnett based on 
$\varepsilon_{\rm eff}^{\rm MG}$ given by Eq.(\ref{MGeff}).

Regarding the EEL spectrum in the dilute photonic crystal, our effective 
dielectric function as well as the Maxwell-Garnett approximation  
reproduce the spectrum having a single peak near $\omega=\omega_p/\sqrt{2}$
fairly well.

\section{Summary}
In this paper we have presented a fully-relativistic analysis of the 
EEL and the induced radiation emission in various spatial arrangements of 
metallic cylinders 
by using the multiple scattering method and the layer-KKRO method. 
In an isolated metallic cylinder with a nanoscale diameter  
we showed that the EEL is dominated by the absorption rather than the 
induced radiation emission.  Thus, the efficiency of converting the kinetic 
energy of the charged particle to the radiation emission is very low. 
In the two identical metallic cylinders a variety of EEL peaks appear. 
Some of them are attributed to the cavity mode localized in the groove 
between the cylinders.  Such a cavity mode as well as the SPP modes 
become the seed of the flat bands 
in a dense periodic arrangement of the metallic cylinders. 
After presenting a mathematical description of the EEL and the SP radiation 
emission 
in two-dimensional photonic crystals composed of cylinders, 
we showed the numerical results of the EEL spectra in both dilute and 
dense periodic arrays of the 
metallic cylinders. In the dilute photonic crystal the EEL spectrum has 
a simple structure. 
The spectrum has a single peak near $\omega=\omega_p/\sqrt{2}$ and 
is not so far from the EEL spectrum in the isolated cylinder. 
However, when a high speed charged particle passes near the photonic crystal, 
a sequence of very sharp loss peaks, which comes from the lowest photonic band 
guided in the finite-thick photonic crystal, are observed. 
The peaks are comparable in magnitude with that by the SPP bands.  
On the other hand in the dense photonic crystal the EEL spectrum is 
very complicated reflecting the photonic band structure, though a good 
correspondence to the EEL spectrum in the almost touched two cylinders 
is observed.  
In both the photonic crystals 
the effective dielectric functions, which are obtained with the reflectance 
of the corresponding semi-infinite photonic crystals, 
fairly reproduce the EEL spectra when   
the charged particle runs inside the photonic crystals.

In this paper we have restricted ourselves to various arrays of Aluminum
cylinders with a diameter of a few nanometers, 
bearing carbon nano-tube arrays in mind. 
Since the plasma wavelength of Aluminum is much larger than the above scale, 
a metallic photonic crystal composed of the cylinders 
behaves as if it has an effective dielectric 
function reflecting the coupled SPPs, in the frequency range concerned. 
As a result, the SP radiation from the photonic crystal is completely absent. 
However, it is of great importance to study the EEL and the 
SP radiation emission spectra when the lattice constant is comparable with 
or larger than the plasma wavelength 
of the constituent cylinders.  In this case the flat bands of coupled SPPs 
appear in $\omega a/2\pi c \simeq O(1)$, so that the EEL is caused both by 
the absorption and by the SP radiation. How these bands as well as usual 
photonic bands affects these spectra is the main theme of the paper II.

\begin{acknowledgments}
The authors would like to thank J. Inoue and S. Yamaguti of Chiba University 
for useful comments.   
This work was supported by ``Promotion of Science and Technology'' 
from the Ministry of Education, Sports, Culture, Science and Technology 
of Japan.   

\end{acknowledgments}


\begin{references}


\bibitem{nano}
U. Kreibig and M. Vollmer, {\it Optical properties of metal clusters} 
(Springer-Verlag, Berlin, 1995). 
\bibitem{Joan}
J.D. Joannopoulos, R.D. Meade, and J.N. Winn, {\it Photonic Crystals}
(Princeton University Press, Princeton, 1995). 
\bibitem{Sakoda}
K. Sakoda, {\it Optical Properties of Photonic Crystals}, 
(Springer-Verlag, Berlin, 2001).   
\bibitem{Veselago}
V.G. Veselago, Sov. Phys. Usp. {\bf 10},  509 (1968).  
\bibitem{Pendry1}
J. B. Pendry, \prl {\bf 85}, 3966 (2000). 
\bibitem{Shelby}
R.A. Shelby, D.R. Smith, and S. Shultz, {\it Science} {\bf 292}, 77 (2001).
\bibitem{Lucas}
A.A. Lucas, L. Hernard, and Ph. Lambin, \prb {\bf 49}, 2888 (1994).
\bibitem{Ritchie}
R.H. Ritchie and A. Howie, Philos. Mag. A {\bf 58}, 753 (1988).  
\bibitem{Schmeits}
M. Schmeits, \prb {\bf 39}, 7567 (1988).
\bibitem{Rivacoba}
J.M. Pitarke and A. Rivacoba, Surf. Sci. {\bf 377}-{\bf 379}, 294 (1997).
\bibitem{Bertsch}
G.F. Bertsch, H. Esbensen, and B.W. Reed, \prb {\bf 58}, 14031 (1998).
\bibitem{PendryMoreno}
J.B. Pendry and L. Martin-Moreno, \prb {\bf 50}, 5062 (1994).
\bibitem{PendryMac}
J.B. Pendry and A. MacKinnon. \prl {\bf 69}, 2772 (1992).
\bibitem{Abajo1}
F.J. Garc\'{\i}a de Abajo, \prl {\bf 82}, 2776 (1999).
\bibitem{Abajo2}
F.J. Garc\'{\i}a de Abajo, \prb {\bf 60}, 6103 (1999).
\bibitem{Abajo3}
F.J. Garc\'{\i}a de Abajo and A. Howie, \prl {\bf 80}, 5180 (1998).
\bibitem{Abajo4}
F.J. Garc\'{\i}a de Abajo and A. Howie, \prb {\bf 65}, 115418 (2002). 


\bibitem{Vidal1}
F.J. Garc\'{\i}a-Vidal and J.M. Pitarke, Eur. Phys. J. B {\bf 22}, 257 (2001).
\bibitem{Yamamoto}  
N. Yamamoto, K. Araya, and F.J. Garc\'{\i}a de Abajo, \prb {\bf 64}, 
205419 (2001).
\bibitem{SP}  
S.J. Smith and E. M. Purcell, Phys. Rev. {\bf 92}, 1069 (1953).
\bibitem{Abajo5} 
F.J. Garc\'{\i}a de Abajo, \pre {\bf 61}, 5743 (2000). 
\bibitem{OhtakaYama} 
K. Ohtaka and S. Yamaguti, Opt. Spectrosc. {\bf 91}, 477 (2001). 
\bibitem{Yamaguti1}
S. Yamaguti, J. Inoue, and K. Ohtaka, \prb {\bf 66}, 085209 (2002).
\bibitem{Yamaguti2}
S. Yamaguti, J. Inoue, O. Haeberl\'{e} and K. Ohtaka, \prb {\bf 66}, 
195202 (2002). 


\bibitem{Landau} 
L.D. Landau, E.M. Lifshitz, and L.P. Pitaevskii, 
{\it Electrodynamics of Continuous Media} (Pergamon Press, Oxford, 1984). 
\bibitem{MG} 
J.C. Maxwell-Garnett, Philos. Trans. R. Soc. London A {\bf 203}, 385 (1904). 
\bibitem{Pitarke} 
J.M. Pitarke, F.J. Garc\'{\i}a-Vidal, and J.B. Pendry, \prb {\bf 57}, 
15261 (1998).  
\bibitem{Smith} 
D.R. Smith, S. Shultz, P. Markos, and C.M. Soukoulis, \prb {\bf 65}, 195104 
(2002).


\bibitem{OhtakaUeta} 
K. Ohtaka, T. Ueta, and K. Amemiya, \prb {\bf 57}, 2550 (1998).


\bibitem{Inoue1}
M. Inoue and K. Ohtaka, \prb {\bf 26}, 3487 (1982).
\bibitem{Inoue2}
M. Inoue and K. Ohtaka, J. Phys. Soc. Jpn {\bf 52}, 1457 (1983).
\bibitem{Inoue3}
M. Inoue and K. Ohtaka, J. Phys. Soc. Jpn {\bf 52}, 3853 (1983).
\bibitem{Vidal2}
F.J. Garc\'{\i}a-Vidal and J. B. Pendry. \prl {\bf 77}, 1163 (1997). 



\bibitem{Ito} 
T. Ito and K. Sakoda, \prb {\bf 64}, 045117 (2001). 
\bibitem{Ochiai} 
T. Ochiai and J. S\'anchez-Dehesa, \prb {\bf 65}, 245111 (2002). 
\bibitem{Kuzmiak} 
V. Kuzmiak, A.A. Maradudin, and F. Pincemin,  \prb {\bf 50}, 16835 (1994). 
\bibitem{Pendry3} 
J.B. Pendry, A.J. Holden, W.J. Stewart, and I. Youngs, \prl {\bf 76}, 
4773 (1996) 


\bibitem{OhtakaNumata}
K. Ohtaka and H. Numata, Phys. Lett. {\bf 73A}, 411 (1979).
\bibitem{Botten}
L.C. Botten, N.A. Nicorovici, R.C. McPhedran, C. Martijn de Sterke, 
and A.A. Asatryan, \pre {\bf 64}, 046603 (2001).  

\bibitem{Korringa}
J. Korringa, Physica {\bf 13}, 392 (1947). 
\bibitem{Kohn}
W. Kohn and N. Rostoker, Phys. Rev. {\bf 94} 1111 (1954).  
\bibitem{Ohtaka1}
K. Ohtaka, \prb {\bf 19}, 5057 (1979).  
\bibitem{Ohtaka2}
K. Ohtaka, J. Phys. C {\bf 13}, 667 (1980).  
\bibitem{Modinos}
A. Modinos, Physica A {\bf 141}, 575 (1987). 
\bibitem{Stefanou}
N. Stefanou, V. Karathanos, and A. Modinos, J. Phys. Condens. Matter {\bf 4}, 
7389 (1992).
\bibitem{Abajo6}
F.J. Garc\'{\i}a de Abajo and L. A. Blanco, \prb {\bf 67}, 125108 (2003). 


\end{references}

\end{document}